\documentclass[conference]{IEEEtran}

\usepackage{graphicx, amsfonts}
\usepackage{epstopdf}
\usepackage[fleqn]{amsmath}
\usepackage{amssymb}
\usepackage{algorithmicx}
\usepackage[ruled]{algorithm}
\usepackage[noend]{algpseudocode}
\alglanguage{pseudocode}

\usepackage{threeparttable}
\usepackage{cite}

\usepackage{caption}
\usepackage{subfig}
\usepackage{float}
\usepackage{graphicx}
\usepackage{tablefootnote}
\usepackage{booktabs}
\usepackage{multirow}
\usepackage[T1]{fontenc}
\usepackage{pgf}
\usepackage[normalem]{ulem}
\usepackage[autoplay]{animate}

\usepackage{amsmath}
\usepackage{fixmath}
\usepackage{amsfonts}
\definecolor{coloShiftSRC}{rgb}{0.36, 0.54, 0.66}

\definecolor{colorlcksvd}{rgb}{0.91, 0.84, 0.42}
\definecolor{colorlcksvdd}{rgb}{0.8, 0.0, 0.1}
\definecolor{colorlcksvd}{rgb}{1, 0.56, 0.0}
\definecolor{colorfddl}{rgb}{0.44, 0.16, 0.39}
\definecolor{colordlsi}{rgb}{0.55, 0.71, 0.0}
\definecolor{colorcopar}{rgb}{0.9, .0, 0}
\definecolor{darkcyan}{rgb}{0.0, 0.55, 0.55}
\definecolor{colordlr}{rgb}{0.0, 0.55, 0.55}
\definecolor{colorlrsdl}{rgb}{0.0, 0.2, 1.0} 
\definecolor{colorlck}{rgb}{0.0, 0.9, 0.9}
\definecolor{pinegreen}{rgb}{0.0, 0.47, 0.44}

\def\L{{\cal L}}

\def\bmt{\left[\begin{matrix}}

\def\emt{\end{matrix}\right]}

\def\bx{\mathbf{x}}
\def\ba{\mathbf{a}}

\def\bg{\mathbf{g}}

\def\by{\mathbf{y}}

\def\and{\text{~and~}}

\def\bD{\mathbf{D}}

\def\bX{\mathbf{X}}

\def\bbX{\lbar{\bX}}        
\def\bbx{\lbar{\bx}}

\makeatletter
\newsavebox\myboxA
\newsavebox\myboxB
\newlength\mylenA
\newcommand*\lbar[2][.75]{%
    \sbox{\myboxA}{$\m@th#2$}%
    \setbox\myboxB\null
    \ht\myboxB=\ht\myboxA%
    \dp\myboxB=\dp\myboxA%
    \wd\myboxB=#1\wd\myboxA
    \sbox\myboxB{$\m@th\overline{\copy\myboxB}$}
    \setlength\mylenA{\the\wd\myboxA}
    \addtolength\mylenA{-\the\wd\myboxB}%
    \ifdim\wd\myboxB<\wd\myboxA%
       \rlap{\hskip 0.5\mylenA\usebox\myboxB}{\usebox\myboxA}%
    \else
        \hskip -0.3\mylenA\rlap{\usebox\myboxA}{\hskip 0.3\mylenA\usebox\myboxB}%
    \fi}
\makeatother

\def\L{\mathcal{L}}       
\ifCLASSINFOpdf
\else
\fi
\begin{document}
%

\title{{Deep Network for Simultaneous Decomposition and Classification in UWB-SAR Imagery}}
\author{Tiep H. Vu$^1$,  Lam Nguyen$^2$, Tiantong Guo$^3$,Vishal Monga$^4$ \\
\begin{tabular}[t]{c@{\extracolsep{4em}}c} 
 \\ $^{1, 3, 4}$ School of Electrical Engineering and Computer Science  & $^{2}$ RF Signal Processing and Modeling Branch\\
The Pennsylvania State University & U.S. Army Research Laboratory \\ 
University Park, PA 16802 & 2800 Powder Mill Rd, Adelphi, MD 20783 \\
Email: $^1$ tiepvu@psu.edu  & Email: $^2$ lam.h.nguyen2.civ@mail.mil \\
$^3$ txg211@psu.edu &  \\
$^4$ vmonga@engr.psu.edu & 
\end{tabular}
}

\maketitle

\begin{abstract}

Classifying buried and obscured targets of interest from other natural and manmade clutter objects in the scene is an important problem for the U.S. Army. Targets of interest are often represented by signals captured using low-frequency (UHF to L-band) ultra-wideband (UWB) synthetic aperture radar (SAR) technology. This technology has been used in various applications, including ground penetration and sensing-through-the-wall. However, the technology still faces a significant issue regarding low-resolution SAR imagery in this particular frequency band, low radar cross sections (RCS), small objects compared to radar signal wavelengths, and heavy interference. The classification problem has been firstly, and partially, addressed by sparse representation-based classification (SRC) method which can extract noise from signals and exploit the cross-channel information. Despite providing potential results, SRC-related methods have drawbacks in representing nonlinear relations and dealing with larger training sets. In this paper, we propose a Simultaneous Decomposition and Classification Network (SDCN) to alleviate noise inferences and enhance classification accuracy. The network contains two jointly trained sub-networks: the decomposition sub-network handles denoising, while the classification sub-network discriminates targets from confusers. Experimental results show significant improvements over a network without decomposition and SRC-related methods.

\end{abstract} 


%
\IEEEpeerreviewmaketitle

\section{Introduction}

\label{sec:intro}

Over the past two decades, the U.S. Army has been investigating the capability of low-frequency, ultra-wideband (UWB) synthetic aperture radar (SAR) systems.
These systems are especially suitable for the detection of buried and obscured targets in various applications, such as foliage penetration~\cite{lam1997}, ground penetration~\cite{lam1998}, and sensing-through-the-wall~\cite{lam2008}. To achieve both resolution and penetration capability, these systems must operate in the low-frequency spectrum, which spans the
 UHF frequency band to L band. Although much progress has been made over the years, one critical challenge still facing the low-frequency UWB SAR technology is the discrimination of targets of interest from other natural and manmade clutter objects in the scene. The key issue of this problem is that the targets of interest are typically small compared to the wavelengths of the radar signals in this frequency band and have very low radar cross sections (RCSs). Thus, it is very difficult to discriminate targets and clutter objects using low-resolution SAR imagery.

\begin{figure*}[t]
\centering
\includegraphics[width = .99\textwidth]{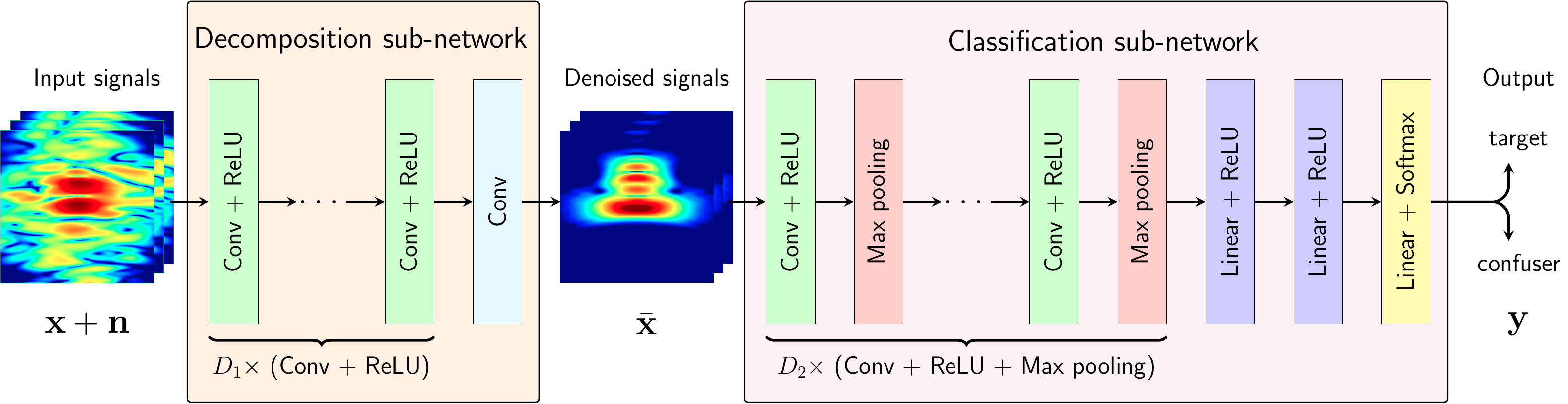}
\caption{\small The proposed Simultaneous Decomposition and Classification Network. The Decomposition Network plays the role of decomposing noisy signals into clean and noise parts. Since the noisy signal does not contribute to the classification process, only latent clean signals $\bbx$ are maintained. The Classification Network predicts label of signals based on the latent clean signals.}
\label{fig:SDCN}
\end{figure*}


The problem has been successfully tackled in previous work~\cite{vu2017tensor} using various sparse representation-based classification (SRC) frameworks. In that paper, the traditional SRC~\cite{Wright2009SRC} framework was generalized to models that can exploit information from a shared class, e.g., background. These models also are capable of handling multichannel classification problems using structures of sparse coefficients using various techniques. The central idea of these models is twofold. {\it First}, forcing tensor sparsity among all channels enhances classification performance significantly. {\it Second}, highly corrupted signals are represented by a conjunction of two parts: ground signals and signals of interest (targets or confusers). Ground signals are common for all classes and can be extracted using a ``shared dictionary''which comprises all training ground signals. The signals of interest are more discriminative, and therefore, more useful for classification purposes. One key observation is that once ground signals are eliminated, classification accuracy can reach approximately 95 \%. In spite of obtaining high accuracy in cases of {\it mostly clean} signals, SRC-related frameworks still struggle with more realistic signals when objects are highly corrupted due to being buried under extremely {\it rough} ground. In addition, when the training set becomes larger, those methods also suffer from a high computational burden at test time, even if dictionary learning methods~\cite{vu2016learning,vu2016tmi,vu2016fast} are incorporated to compact the dictionary. Furthermore, the core idea of SRC is highly based on the linearity of the signals, which is unrealistic in real battlefields.

Compared to SRC-related methods, deep learning~\cite{lecun2015deep} methods have been proven to provide better results in numerous machine learning and signal processing problems, especially when more training data were involved. Particularly, denoising~\cite{bengio2009learning,Zhang2017} and classification~\cite{krizhevsky2012imagenet,he2016deep} problems have received much attention recently, thanks to the powerful representation capability of deep networks. Additionally, deep learning is considered the state of the art in the problems where the mapping function between input and output is unknown or highly nonlinear. For classification tasks, deep networks often contain many convolutional layers in conjunction with various kind of layers, e.g., max pooling, batch normalization~\cite{ioffe2015batch}, dropout~\cite{srivastava2014dropout}, etc., for particular problems. 

Being mindful of the importance and challenges of our problem, we aim to build a convolutional network that is capable of decomposition and classification of signals simultaneously, enhancing the desired overall accuracy. {\bf Our main contribution}: we proposed a convolutional neural network comprising two sub-networks with different roles: decomposition and classification. The decomposition part eliminates noise (ground signals in our case), while the classification part learns to classify the denoised signals as targets or confusers. More importantly, both sub-networks in  the proposed Simultaneous Decomposition and Classification Network (SDCN) are trained at the same time as an end-to-end model. By doing so, each sub-network is collaboratively supported by the other to obtain the common goal -- more accurate classification results. 

\par



\section{Simultaneous Decomposition and Classification Network (SDCN)} 
\label{sec:simultaneous_decomposition_and_classification_network }

\subsection{Previous SRC-related works} 
In~\cite{vu2017tensor}, training signals are combined into a single dictionary $\bD = [\bD_t, \bD_c, \bD_g]$ where $\bD_t, \bD_c, \bD_g$ are sub-dictionaries comprising training signals of {\it targets, confusers} and {\it grounds}. A new noisy signal $\tilde{\bx}$ can be represented by a linear combination of some columns in $\bD$
\begin{eqnarray}
    \tilde{\bx} = \bmt \bD_t & \bD_c & \bD_g \emt 
    \bmt \ba_t \\ \ba_c \\ \ba_g \emt 
\end{eqnarray}
where the coefficient vectors $[\ba_t, \ba_c, \ba_g]^T$ are forced to be sparse by different sparsity constraints. This model can also be extended to tensor cases where a signal is formed by more than one polarizations. 

The ground signals $\bD_g$ could be considered shared information since both noisy targets and confusers can be interfered with the same ground, i.e., buried in the same grounds. With the presence of the shared class, a new test sample is represented by samples from the corresponding class in conjunction with samples from $\bD_g$. Based on this assumption, the identity of $\tilde{\bx}$ is determined by
\begin{equation}
\label{eqn:src}
    \text{identity}(\tilde{\bx}) = \arg\min_{i \in \{t, c\}} \|\underbrace{\tilde{\bx} - \bD_g\ba_g}_{\bbx}
     - \bD_i \ba_i\|_2
\end{equation}
where $(i = t)/(i = c)$ means the signal is classified as a target/confuser. In \eqref{eqn:src}, $\bar{\bx} = \tilde{\bx} - \bD_g\ba_g$ represents the ground elimination process. Accordingly, $\min_{i \in \{t, c\}} \|\bar{\bx} - \bD_i\ba_i\|_2$ can be seen as the process of finding which class contributes more to the reconstruction of the input noisy signal $\tilde{\bx}$.

The success of SRC-related methods is based on the key idea of the separated dictionary $\bD_g$. This technique is somewhat difficult to formulate in a neural network-based method. However, information about the ground can still be obtained in deep networks with many noisy training signals, and from those, information about the noise can be learned and extracted to enhance classification accuracy. A noisy input signal can be generated by randomly selecting a sample $\bx$ in $[\bD_t, \bD_c]$ (an object), a ground sample $\bx_g$ in $\bD_g$, and a random positive number $\lambda$, then finally
$$\tilde{\bx} = \bx + \lambda \bg$$

Although number of $\bx, \bg$ is limited, the random real number $\lambda$ can be chosen unlimitedly to form several input training samples $\tilde{\bx}$. The corresponding output (label) is simply the label of $\bx$.

\subsection{General flow of signals in SDCN} 
\label{sub:general_flow_of_signal_in_sdcn}
The flow of signals in SDCN is visualized in Fig.~\ref{fig:SDCN}.

The input of SDCN is a noisy observation $\tilde{\mathbf{x}} = \mathbf{x + n}$ and is a combination of $\mathbf{x}$, the clean signal captured at ideal condition, and $\mathbf{n}$, the representation of the ground without the presence of objects of interest. It is worth noting that this configuration is made for training only where our system is capable of capturing those signal separately. The first part, the Decomposition sub-network, aims to learn a mapping function $g_{\Theta_1}(\tilde{\bx}) = \bbx$ to predict the latent clean signal with $\Theta_1$ being the set of all parameters in this first network. Those parameters are trained to obtain as small a difference between the latent clean signals $\bbx$ and the true clean signals $\mathbf{x}$ as possible. Concretely, the averaged mean squared error between these signals is forced to be small:
\begin{eqnarray}
\label{eqn:decompose_loss}
    \arg\min_{\Theta_1}\mathcal{L}_1(\Theta_1) = \frac{1}{2N}\sum_{i=1}^{N}\|\bbx_i - \bx_i\|_2^2 = \frac{1}{2N} \|\bbX - \bX\|_F^2
\end{eqnarray}
where $N$ is the total number of training signals, $\bX = [\bx_1, \dots, \bx_N]$ and $\bbX = [\bbx_1, \dots, \bbx_N]$.

After getting $\bbx$, SDCN continues feeding this approximated clean signal into its second part - the Classification sub-network. The Classification sub-network takes each input $\bbx$ and generates a probability vector $\by = h_{\Theta_2}(\bbx)$. Each element in $\by$ is the estimated probability of the original signal belonging to each corresponding class. In our problem, $\by$ has length 2 with the first element representing the {\it target} and the other being the {\it confuser}. To maximize the performance, the output vector $\by$ is forced to be close to the one-hot vector $\hat{\by}$ representing the true label of the input signal. Formally, the objective function of the second network is formulated as: 
\begin{eqnarray}
\label{eqn:classification_loss}
    \mathcal{L}_2(\Theta_2) = \frac{1}{N}\sum_{i=1}^N H(\hat{\by}, \by)
\end{eqnarray}
where $\Theta_2$ is the set of all parameters in the sub-network and $H(\by, \hat{\by})$ is the cross-entropy function of two probability vectors $\hat{\by}$ and $\by$ with $C$ classes:
\begin{eqnarray}
    H(\hat{\by}, \by) = -\sum_{c=1}^C \hat{y}_i \log(y_i)
\end{eqnarray}

In fact, the training process of the ultimate classification problem could be divided into two parts. In the first part, the Decomposition sub-network is trained individually. After being completely trained, the first sub-network can be seen as a feature generator to create training samples for the Classification sub-network. The second sub-network is trained on the latent clean signals $\bbx$. In the test process, a noisy input $\bx + \mathbf{n}$ is first fed into the Decomposition sub-network. Its predicted clean signal $\bbx$ is then put into the Classification sub-network to obtain the predicted label. 

However, it has been proved that training an end-to-end model often produces 
better results compared to the above two-step framework. In other words, both 
of the above sub-networks could be train simultaneously. By doing so, the 
Decomposition sub-network not only learns to {\it clean} signals but also to 
keep discriminative information that is useful for the Classification 
sub-network. Both sub-networks are combined into one and can be trained at the
same time by combining their loss functions. Straightforwardly, the overall loss function of the whole network is the weighted sum of the decomposition 
loss in \eqref{eqn:decompose_loss} and the classification loss in 
\eqref{eqn:classification_loss}:
\begin{equation}
\label{eqn:mainloss}
    \L_{\text{SDCN}}(\Theta_1, \Theta_2) = \frac{1}{2N} \|\bbX - \bX\|_F^2 + \gamma \frac{1}{N}\sum_{i=1}^NH(\hat{\by}, \by)
\end{equation}
where $\gamma$ is a positive regularization parameter balancing the importance of each sub-network. By minimizing this loss function, we can jointly obtain a model for both signal decomposition and classification.

\subsection{Network architecture} 
\label{sub:network_architecture}
\subsubsection{Decomposition sub-network}
The proposed Decomposition sub-network comprises $D_1$ (Conv + ReLU) layers and a convolutional layer at the end. Each Conv+ReLU contains 64 filters of size $3 \times 3 \times c$ for generating 64 feature maps and one rectified linear unit (ReLU) for simple nonlinearity. In our setup, $c = 64$ for every layer except the first layer, where $c = 1, 2,$ or 3, depending on the number of polarizations used. The last layer in this sub-network does not include the ReLU since the desired clean signals can be negative. For every convolution unit in this sub-network, the zero padding technique  is used to assure that $\bbx$ and $\bx$ have the same sizes. 

\subsubsection{Classification sub-network}
There are four different types of layers shown in the right part of Fig.~\ref{fig:SDCN} with different colors:

\begin{itemize}
    \item $D_2$ Conv+ReLU layers: these are similar to the Conv+ReLU in the Decomposition sub-network except that the zero padding is negligible at the boundary. The zero padding could be omitted since the output does not require same size as the input, but is a 2-D probability vector.

    \item $D_2$ Max pooling layers: these ($2\times 2$) max pooling layers play an important role in the classification problem in general. Max pooling layers not only reduce feature map dimensions after each layer but also have been proven to generate features that are invariant to translation. 

    \item Two Linear+ReLU layers: the first Linear + ReLU layer simply converts features from the tensor/matrix form to a vector of dimension 512. The second Linear + ReLU layer continues to generate lower-dimensional (128) features. These two layers are also called fully connected layers.

    \item One Linear + Softmax layer: this is another fully connected layer where the activation function is softmax instead of ReLU. The softmax generates a 2-D vector whose elements are positive and summed up to one. The final 2-D vector can be considered the probability vector. 

\end{itemize}

\begin{figure}
  \includegraphics[width = .49\textwidth]{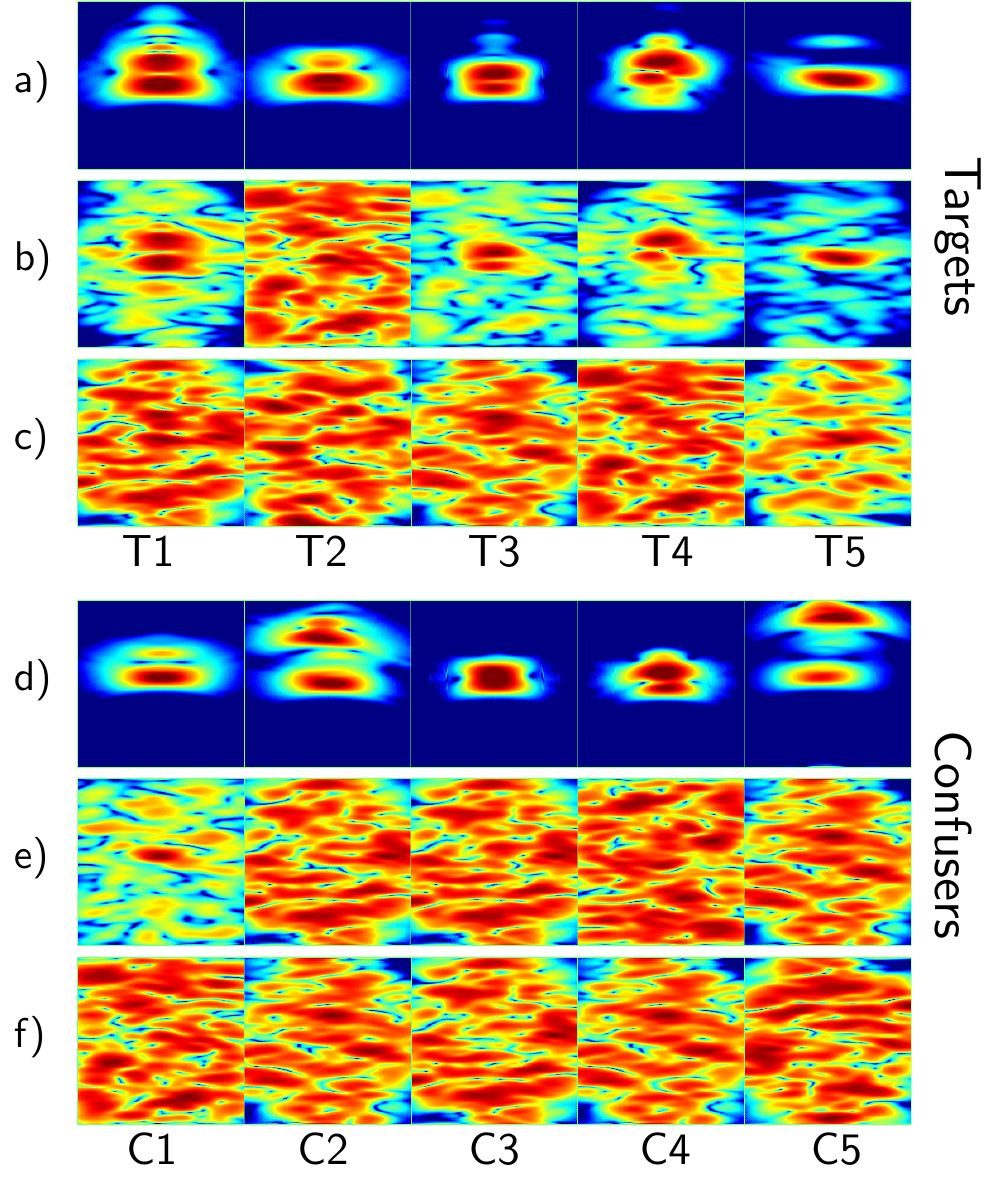}
\caption{\small Sample images of five targets and five clutter objects. T1 = M15 anti-tank mine, T2 = TM62P3 plastic mine, T5 = 155 mm artillery shell, C1 = coke can, C2 = rocks, C3 = rocks, C4 = rocks, C5 = rocks. a) Targets under smooth ground surface. b) Targets under rough ground surface (easy case, scale = 1). c) Targets under rough ground surface (hard case, scale = 5). d) Confusers under smooth ground surface. e) Confusers under rough ground surface (easy case, scale=1). f) Confusers under rough ground surface (hard case, scale=5).}
\label{fig:examples_target}
\end{figure}


\section{Experimental results} 
\label{sec:experimental_results}







\begin{figure}[t]
\centering
\includegraphics[width = .48\textwidth]{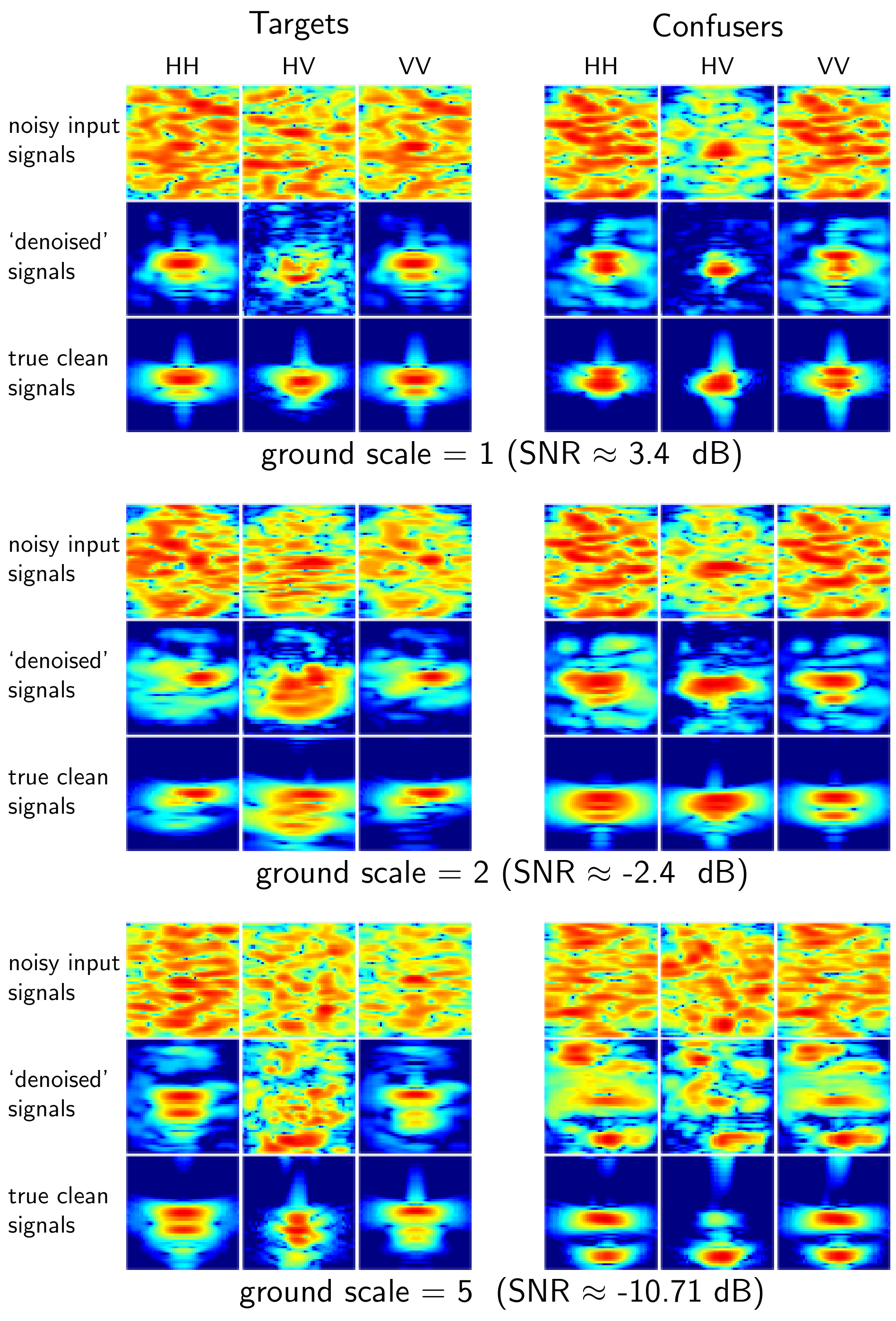}
\caption{\small Visualization of signals after training SDCN. Left column - targets, right column - confusers. Row 1 -- ground scale = 1 (easy), row 2 -- ground scale = 2 (less easy), row 3 -- ground scale = 5 (very difficult). In each $3\times 3$ block: top row -- noisy input signals ($\tilde{\bx}$), middle row -- denoised signals ($\bbx$), bottom row -- ground truth ($\bx$), left column -- horizontal transmitter, horizontal receiver (HH), middle column -- horizontal transmitter, vertical receiver (HV), right column -- vertical transmitter, vertical receiver (VV).}
\label{fig:denoise_results}
\end{figure}

\begin{figure*}[t]
\centering
\includegraphics[width = .99\textwidth]{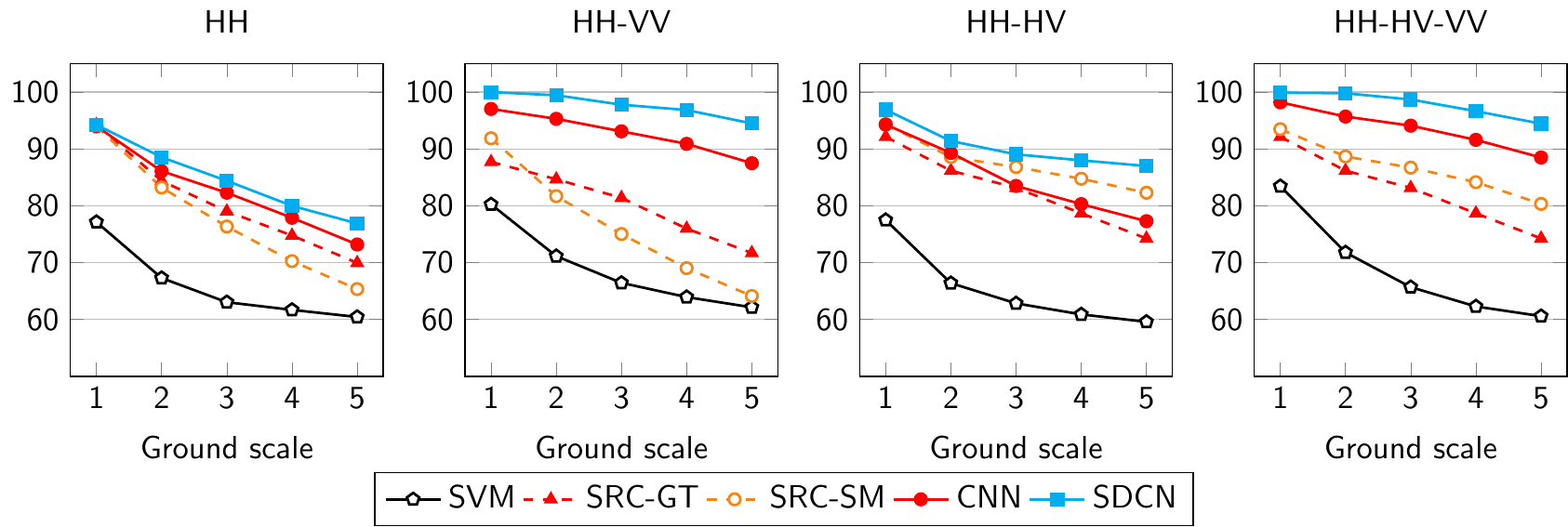}
\caption{\small Classification accuracy (\%) as a function of number of noise levels and polarization combinations.}
\label{fig:pol_combine_test}
\end{figure*}

\begin{figure}[t]
\centering
\includegraphics[width = .48\textwidth]{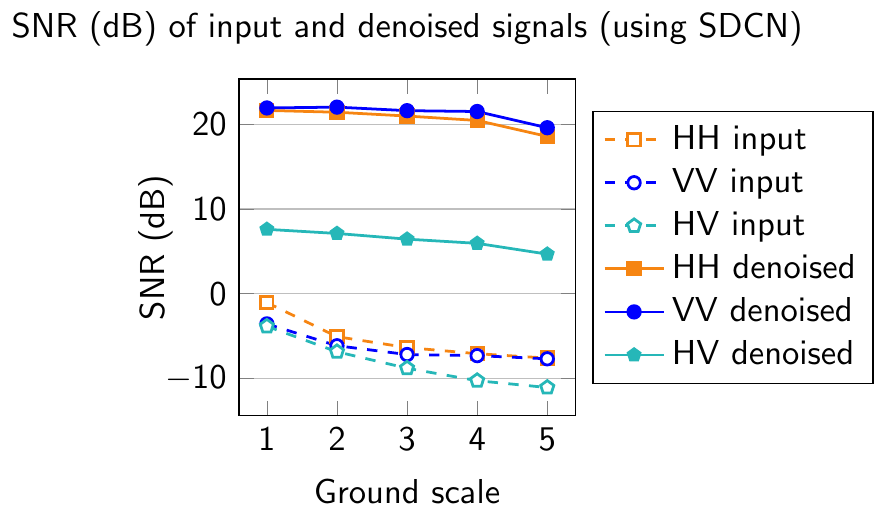}
\caption{\small Signal-to-noise ratio (SNR) of input signals and denoised signals obtained at the output of the SDCN sub-network.}
\label{fig:snr_results}
\end{figure}

\subsection{Original Dataset}

The proposed SDCN is applied to a simulated SAR database consisting of targets (metal and plastic mines, 155-mm unexploded ordinance [UXO], etc.) and clutter objects (soda cans, rocks, etc.) buried under rough ground surfaces. The electromagnetic (EM) radar data are simulated based on the full-wave computational EM method known as finite-difference, time-domain (FDTD) software~\cite{dogaru2010}, which was developed by the U.S. Army Research Laboratory (ARL). The software was validated for a wide variety of radar signature calculation scenarios~\cite{dogaru2007,liao2012}.  
SAR images are formed using backprojection image formation~\cite{McCorkle1994}
with an integration angle of $30^{\circ}$ (please refer to~\cite{vu2017tensor}
for more details). Fig.~\ref{fig:examples_target}a shows  SAR images (using
vertical transmitter, vertical receiver -- VV -- polarization) of some targets
that are buried under a perfectly smooth ground surface.
Fig.~\ref{fig:examples_target}b and~\ref{fig:examples_target}c shows the same
targets as Fig.~\ref{fig:examples_target}a, except that they are buried under
rough ground surfaces (the easiest case  corresponds to ground scale = 1 and the
harder case corresponds to ground scale = 5). Similarly,
Fig.~\ref{fig:examples_target}d,~\ref{fig:examples_target}e,
and~\ref{fig:examples_target}f show the SAR images of some clutter objects
buried under a smooth and rough surfaces.

 For training, the target and clutter objects are buried under a smooth surface to generate high signal-to-clutter ratio images. We include 12 SAR images that correspond to 12 different aspect angles ($0^{\circ}, 30^{\circ},\dots,330^{\circ}$) for each target type. To incorporate the ground information into training, we also generate corresponding rough ground images without presence of objects of interest. 
 For testing, SAR images of targets and confusers are generated at 100 random aspect angles and buried under rough ground surfaces. 
Various levels of ground surface roughness are simulated by selecting different ground surface scaling factors when embedding the test targets under the rough surfaces. 


\subsection{Experimental setup} 
\label{sub:}

There are five different methods considered in our problem: support vector machine with the radial basic function kernel (SVM), SRC~\cite{vu2017tensor} with simultaneous constraint (SRC-SM), SRC with group tensor sparsity constraint (SRC-GT), a general convolution neural network for classification (CNN), and the proposed SDCN. In our experiment, CNN and SDCN have the same network structures, but the loss function of CNN contains the cross-entropy term only. In other words, CNN ignores the importance of the decomposition part. For SDCN with loss function in \eqref{eqn:mainloss}, we simply set the regularization parameter $\gamma = 1$. The depths of the decomposition sub-network and the classification sub-network are $D_1 = 10$ and $D_2 = 3$.

For the two SRC-based frameworks, training samples are used to form a big dictionary $\bD = [\bD_t, \bD_c, \bD_g]$ where $\bD_t, \bD_c, \bD_g$ are sets of {\it target samples, confuser samples}, and {\it ground samples} respectively. By including $\bD_g$ in the dictionary, both frameworks also have ability to decompose the input signals into two parts: clean signals corresponding to $[\bD_t, \bD_c]$ and the ground represented by $\bD_g$. It is noted that each dictionary can be a tensor if more than one polarization is used in the experiment. 

For SVM and the two deep learning-based frameworks, training samples are generated by
\begin{eqnarray}
\label{eqn:mixdata}
  \tilde{\bx} = \bx + \lambda \bg  
\end{eqnarray}
where $\bx$ is randomly sampled from the either {\it targets} or {\it confusers} training set; $\bg$ is randomly selected from {\it ground} training set; and the level of noise $\lambda$ is uniformly chosen in the range $[0.5, 5.5]$. The wide range is picked to make sure that the training samples cover all noise levels from 1 to 5 in the test samples. For each class, 10,000 samples are generated for training. The label of each training data $\tilde{\bx}$ is set to be identical to the label of the clean signal $\bx$. This process is called data augmentation. In fact, no more training data are used in these methods, since all 20,000 samples are generated from the same original training set. 

Test samples for all methods are created in the same way as \eqref{eqn:mixdata} on the original test data. The levels of noise $\lambda$ are fixed at \{1, 2, 3, 4, 5\} for result reporting purposes. 

{\textbf{Remark:} Among five mentioned methods, SVM and CNN lack the ability of denoising data, while two SRC methods and the proposed SDCN are capable of doing so. }

\subsection{Denoised signals visualization} 
\label{sub:decomposed_result_visualization}
The proposed SDCN is trained and its loss function converges after 50 epochs. To investigate the decomposition ability of the network, we visualize in Fig.~\ref{fig:denoise_results} the latent clean signals with inputs being test data at different levels of noise. 

We can see that SDCN successfully eliminate the ground from the noisy signals; the obtained denoised signals are visually close to the ground truth at ground scales 1 and 2. At the harder level when the ground scale equals 5, the network still performs well. It is also noted that the recovered HV signals are noisier. This effect is predictable since cross polarization HV signals are weaker than those at other polarizations. 
    
{To formally confirm the effect of the Decomposition sub-network, we show the signal-to-noise ratios (SNR) of each polarization input signals and their corresponding denoised signals in Fig.~\ref{fig:snr_results}. It can be seen that SNRs of input signals (dashed lines) are very low with all of them being below zero. In contrast, the denoised signals (dashed line) show that the Decomposition sub-network significantly improve quality of signals with big gains occurring in HH and VV polarizations.}

\subsection{Overall classification accuracy} 
{Classification accuracy (\%) of the five methods on the dataset are shown in Fig.~\ref{fig:pol_combine_test} at different levels of noise and for different polarization combinations. It is clear that the two deep learning methods (solid lines), in general, outperform the others with big gaps in the HH-VV and HH-HV-VV combinations. In addition, SDCN provides better results than CNN with the gap widening as noise level increases. This observation confirms the importance contribution of the Decomposition sub-network.}

{The mostly identical results of neural networks and sparse representation-based methods occur in the HH-HV combination. This can be explained by the fact that cross polarization HV signals worsen the results due to their low signal-to-noise ratio. It also can be seen that in this combination, CNN is outperformed by both SRC-SM and SDCN since it lacks of denoising capability. Similarly, without this important capability, SVM is beaten by all others with noticeable gaps. 
}

\section{Conclusion}

In this paper, we propose a convolutional neural network for the SAR UWB imagery classification problem. To obtain a good result, in addition to the Classification sub-network, the signal processing part is also included in the network performed by the Decomposition sub-network. Both sub-networks are trained simultaneously to obtain one final network, which can successfully classify signals even when the signals are extremely corrupted. The classification results show that neural networks outperform sparse representation-based methods with significant gaps.

\bibliographystyle{IEEEtran}
\bibliography{refs}

\begin{thebibliography}{10}
\providecommand{\url}[1]{#1}
\csname url@samestyle\endcsname
\providecommand{\newblock}{\relax}
\providecommand{\bibinfo}[2]{#2}
\providecommand{\BIBentrySTDinterwordspacing}{\spaceskip=0pt\relax}
\providecommand{\BIBentryALTinterwordstretchfactor}{4}
\providecommand{\BIBentryALTinterwordspacing}{\spaceskip=\fontdimen2\font plus
\BIBentryALTinterwordstretchfactor\fontdimen3\font minus
  \fontdimen4\font\relax}
\providecommand{\BIBforeignlanguage}[2]{{%
\expandafter\ifx\csname l@#1\endcsname\relax
\typeout{** WARNING: IEEEtran.bst: No hyphenation pattern has been}%
\typeout{** loaded for the language `#1'. Using the pattern for}%
\typeout{** the default language instead.}%
\else
\language=\csname l@#1\endcsname
\fi
#2}}
\providecommand{\BIBdecl}{\relax}
\BIBdecl

\bibitem{lam1997}
L.~H. Nguyen, R.~Kapoor, and J.~Sichina, ``Detection algorithms for
  ultrawideband foliage-penetration radar,'' \emph{Proceedings of the SPIE,
  3066}, pp. 165--176, 1997.

\bibitem{lam1998}
L.~H. Nguyen, K.~Kappra, D.~Wong, R.~Kapoor, and J.~Sichina, ``Mine field
  detection algorithm utilizing data from an ultrawideband wide-area
  surveillance radar,'' \emph{Proceedings of the SPIE International Society of
  Optical Engineers, 3392}, no. 627, 1998.

\bibitem{lam2008}
L.~H. Nguyen, M.~Ressler, and J.~Sichina, ``Sensing through the wall imaging
  using the {A}rmy {R}esearch {L}ab ultra-wideband synchronous impulse
  reconstruction ({U}{W}{B} {S}{I}{R}{E}) radar,'' \emph{SPIE}, no. 69470B,
  2008.

\bibitem{vu2017tensor}
T.~H. Vu, L.~Nguyen, C.~Le, and V.~Monga, ``Tensor sparsity for classifying
  low-frequency ultra-wideband ({U}{W}{B}) {S}{A}{R} imagery,'' in \emph{IEEE
  Radar Conference (RadarConf)}, 2017, pp. 0557--0562.

\bibitem{Wright2009SRC}
J.~Wright, A.~Yang, A.~Ganesh, S.~Sastry, and Y.~Ma, ``Robust face recognition
  via sparse representation,'' \emph{IEEE Trans.\ on Pattern Analysis and
  Machine Intelligence}, vol.~31, no.~2, pp. 210--227, Feb. 2009.

\bibitem{vu2016learning}
T.~H. Vu and V.~Monga, ``Learning a low-rank shared dictionary for object
  classification,'' in \emph{Proc.\ IEEE Conf.\ on Image Processing}.\hskip 1em
  plus 0.5em minus 0.4em\relax IEEE, 2016, pp. 4428--4432.

\bibitem{vu2016tmi}
T.~H. Vu, H.~S. Mousavi, V.~Monga, U.~Rao, and G.~Rao, ``Histopathological
  image classification using discriminative feature-oriented dictionary
  learning,'' \emph{IEEE Transactions on Medical Imaging}, vol.~35, no.~3, pp.
  738--751, March, 2016.

\bibitem{vu2016fast}
T.~H. Vu and V.~Monga, ``Fast low-rank shared dictionary learning for image
  classification,'' \emph{IEEE Transactions on Image Processing}, vol.~26,
  no.~11, pp. 5160--5175, Nov 2017.

\bibitem{lecun2015deep}
Y.~LeCun, Y.~Bengio, and G.~Hinton, ``Deep learning,'' \emph{Nature}, vol. 521,
  no. 7553, pp. 436--444, 2015.

\bibitem{bengio2009learning}
Y.~Bengio \emph{et~al.}, ``Learning deep architectures for {A}{I},''
  \emph{Foundations and trends{\textregistered} in Machine Learning}, vol.~2,
  no.~1, pp. 1--127, 2009.

\bibitem{Zhang2017}
K.~Zhang, W.~Zuo, Y.~Chen, D.~Meng, and L.~Zhang, ``Beyond a {G}aussian
  denoiser: Residual learning of deep {C}{N}{N} for image denoising,''
  \emph{IEEE Transactions on Image Processing}, vol.~7, pp. 3142--3155, 2017.

\bibitem{krizhevsky2012imagenet}
A.~Krizhevsky, I.~Sutskever, and G.~E. Hinton, ``Image{N}et classification with
  deep convolutional neural networks,'' in \emph{Advances in Neural Information
  Processing Systems}, 2012, pp. 1097--1105.

\bibitem{he2016deep}
K.~He, X.~Zhang, S.~Ren, and J.~Sun, ``Deep residual learning for image
  recognition,'' in \emph{IEEE {C}onference on {C}omputer {V}ision and
  {P}attern {R}ecognition}, 2016, pp. 770--778.

\bibitem{ioffe2015batch}
S.~Ioffe and C.~Szegedy, ``Batch normalization: Accelerating deep network
  training by reducing internal covariate shift,'' in \emph{International
  Conference on Machine Learning}, vol.~37, 2015, pp. 448--456.

\bibitem{srivastava2014dropout}
N.~Srivastava, G.~E. Hinton, A.~Krizhevsky, I.~Sutskever, and R.~Salakhutdinov,
  ``Dropout: a simple way to prevent neural networks from overfitting.''
  \emph{Journal of Machine Learning Research}, vol.~15, no.~1, pp. 1929--1958,
  2014.

\bibitem{dogaru2010}
T.~Dogaru, ``A{F}{D}{T}{D} user's manual,'' \emph{ARL Technical Report,
  Adelphi, MD, ARL-TR-5145}, March 2010.

\bibitem{dogaru2007}
T.~Dogaru, L.~Nguyen, and C.~Le, ``Computer models of the human body signature
  for sensing through the wall radar applications,'' \emph{ARL, Adelphi, MD,
  ARL-TR-4290}, September 2007.

\bibitem{liao2012}
D.~Liao and T.~Dogaru, ``Full-wave characterization of rough terrain surface
  scattering for forward-looking radar applications,'' \emph{IEEE Trans. on
  Antenna and Propagation}, vol.~60, pp. 3853--3866, August 2012.

\bibitem{McCorkle1994}
J.~McCorkle and L.~Nguyen, ``Focusing of dispersive targets using synthetic
  aperture radar,'' \emph{U.S. Army Research Laboratory, Adelphi, MD,
  ARL-TR-305}, August 1994.

\end{thebibliography}
%




\end{document}